\documentstyle[preprint,aps]{revtex}
\begin{document}

\begin{center}
{\Large \bf Testable Consequences\\
of Curved-Spacetime Renormalization}\\
\bigskip
{Leonard Parker}\\
{Department of Physics}\\
{University of Wisconsin-Milwaukee}\\
{Milwaukee, WI 53201}\\
\end{center}

\bigskip
\begin{abstract}
\noindent 
I consider certain renormalization effects in curved
spacetime quantum field theory.
In the very early universe these effects resemble those of a
cosmological constant, while in the present universe they give
rise to a significant finite renormalization of the gravitational 
constant. The relevant renormalization term
and its relation to elementary particle masses was first found
by Parker and Toms in 1985, as a consequence of
the ``new partially summed form'' of the propagator in curved 
spacetime. The significance of the term is based on
the contribution of massive particles to the vacuum. In the present
universe, this renormalization term appears to account for a large
part or even all of the Newtonian gravitational constant. 
This conjecture is testable because it relates the value of 
Newton's constant to the elementary particle masses.

\end{abstract}

\bigskip
\section{Introduction}
The observational determination of cosmological parameters 
appears to have reached a new level of accuracy. At the
same time, there are longstanding unsettled questions such as the
origin of dark-matter and large-scale structure, as well as
important new questions. Among these is the possibility of a
significant non-zero cosmological constant implied by observations
of type Ia supernovae by two independent groups.\cite{sn1} 
Because of these new observations, I reconsidered a relatively
large renormalization term that arises as a consequence of
general relativity and quantum field theory in curved 
spacetime,\cite{pt1} in the hope that it could explain the non-zero
cosmological constant. This was a particularly appealing possibility
because the relevant term arises through renormalization of the
cosmological constant. Although this term does have significant and
potentially observable consequences, and does act like a 
cosmological constant at very early times, it appears that at the 
present time its main effect is to produce a large finite renormalization
of the gravitational constant. Although it does not seem capable
of producing the acceleration of the expansion of the universe that
is implied by the supernovae observations, it nevertheless has 
consequences that are testable.

In 1985, Parker and Toms\cite{pt1} evaluated the
gravitational part of the quantum corrections to the general 
relativistic effective action for a very general set of
elementary particle quantum fields in an arbitrary curved spacetime.
They showed that if there are particles of high mass contributing
to the vacuum energy, as is predicted by elementary particle and string
theories, then there would be a significant
renormalization of the cosmological constant term in the Einstein-Hilbert
action, and that the magnitude of the renormalization term would 
be of importance even in the present universe.
Here, I consider more fully the role played by this renormalization
term. 

The relevant calculation in Ref.\cite{pt1} is a 
direct evaluation of the gravitational part of  
the effective action that arises from vacuum fluctuations
of quantum fields. The renormalization terms in the effective 
action that are of interest here come from non-perturbative (also
called non-local) \cite{nl} contributions to the propagators of 
quantum fields in curved spacetime. The non-perturbative contribution 
comes from summing, for arbitrary dimensional spacetimes, 
an infinite series consisting 
of {\em all} terms that contain any factors of the scalar 
curvature $R$ in the ``proper-time'' or Schwinger-DeWitt expansion
of the propagator in curved spacetime\cite{pt2,jp1}.
The coefficients in the proper-time expansion contain successively
higher powers of the Riemann tensor, and its covariant derivatives
and contractions.  
The ``new partially summed form of the propagator''
that results from summing all terms containing factors of $R$
is manifestly generally covariant and
contains a non-perturbative exponential factor involving
the scalar curvature, multiplied by a proper-time 
series of terms having (in arbitrary dimension) no factors of $R$.
When the non-perturbative exponential factor is expanded and
the two series are multiplied one recovers the original expansion.
The resulting terms that involve $R$ are quite complicated, and
are not known beyond a few orders.
When the theorem stating that {\em all} of the $R$ terms are
summed in the new partially summed form of the propagator
was first conjectured and proved \cite{pt2,jp1}, only the
first three terms in the series expansion of the propagator  
had been calculated (because of the technical difficulty of 
calculating such complicated expressions). Since that time,
several more terms have been calculated and shown to be
consistent with the new form of the propagator.
It is precisely because the exponential factor in the proper-time
series for the propagator is extracted through the summation
of the {\em complete} infinite series of
highly complicated terms that involve $R$ that
its physical consequences must be regarded as realistic 
predictions of general relativity and quantum field theory.
Although there may be other non-perturbative contributions
to the propagator, this exponential one involving $R$ 
is particularly significant because it gives rise to all the
renormalization effects that involve $R$ in the gravitational
part of the effective action.

I first summarize the result derived in \cite{pt1}
for the renormalized cosmological constant. 
Then I discuss briefly its implications in the very early universe 
and at more length its implications in the present universe. 

\section{Renormalized Cosmological/Gravitational Constant}
In the discussion of the cosmological constant and 
the Newtonian gravitational constant
in Ref.\cite{pt1}, the particle content of the theory under
consideration was quite general, as already noted.
The part of the effective action relevant to our discussion
was found to have the form [see Eq.(3.54) of Ref.\cite{pt1}],
\begin{equation}
\Gamma = \int dv_x (\Lambda_{\rm eff} + \kappa_{\rm eff}R + \cdots),
\label{eq:action}
\end{equation}  
where the quantum corrections to $\kappa_{\rm eff}$ caused by
particles of mass $m$ were significant only at times when
the curvature was of the order $m^2$ or larger. On the other
hand, the quantum corrections to the term $\Lambda_{\rm eff}$
containing the cosmological constant, as noted in Ref.\cite{pt1},
is still of interest at the present time.
The general form of $\Lambda_{\rm eff}$ is given 
(in units with $\hbar = c = 1$) in Eq.(4.5) of Ref.\cite{pt1}:
\begin{equation}
{\Lambda}_{\rm eff} =
\Lambda - \sum_i A_i N_i m_i{}^4\, 
    \ln[ 1 + B_i R/m_i{}^2 ].  \label{eq:lambdasum}
\end{equation}
Here $\Lambda$ is a constant energy-density which must be
determined by observation. Classically, the value of the
cosmological constant $\Lambda_{\rm c}$ is related to the value
of the constant $\Lambda$ in the action by\cite{er1}
$\Lambda_{\rm c} = (8\pi G) \Lambda$
where $G$ is the Newtonian gravitational constant. 
Here the sum is over the various species of particles of
multiplicities $N_i$ and masses $m_i$, and the dimensionless constants
$A_i$ and $B_i$ are of order 1. The sum includes the contributions
from particles of spin-0 and of spin-1/2. The contributions from
particles of spin-1 are somewhat more complicated, but are 
sufficiently illustrated by the terms shown.
For example, for spin-0 particles 
$A_i = (64\pi^2)^{-1}$, and $B_i = \xi_i - 1/6$, where $\xi_i$ is
a dimensionless coupling constant appearing in a term 
of the form $\xi_i R \phi_i{}^2$
in the Lagrangian of the scalar field $\phi_i$. Since the fields
of interest are highly massive, there is no conformal symmetry 
to favor any particular value of $\xi$. Therefore, it is natural 
to assume that $|B_i|$ is of order 1. The sign of the contributions
to the sum in $\Lambda_{\rm eff}$ depends on the sign of $B_i$.

In the early universe, when $R$ is of the order of the particle masses,
the logarithmic terms in $\Lambda_{\rm eff}$ vary slowly and act much
like a cosmological constant. Their effect in the early universe
must be added to those of symmetry breaking scalar fields that
may also contribute to a cosmological constant by being in the 
false vacuum. As the universe expands and $R$ goes from
larger to smaller than the particle masses, one must do
a numerical integration to determine the effect of the 
logarithmic renormalization terms on the evolution. In addition, 
variation of these terms with position and time will affect the 
growth of perturbations at early times.

Because of the factors of $m_i{}^4$, 
the main contribution to the sum in Eq.(\ref{eq:lambdasum})
comes from the largest masses, and except in the very
early universe, the logs containing them can be expanded.
For brevity, 
I will suppress the summation sign and denote by $M$ and $N$ 
the masses $m_i$ and multiplicities $N_i$ of any particle 
types that in recent times satisfy the condition,
$|B_i R/ m_i{}^2| \ll 1$.  To good approximation, in recent times
Eq.(\ref{eq:lambdasum}) becomes
\begin{equation}
\Lambda_{\rm eff} = \Lambda - A B N M^2 R .
\label{eq:lambdasum2}
\end{equation}
Then the effective action of Eq.(\ref{eq:action}) describing the
recent universe becomes,
\begin{equation}
\Gamma = \int dv_x [ \Lambda + (\kappa - A B N M^2) R + \cdots],
\label{eq:action2}
\end{equation}
where terms other than the leading order in $M^2$ have been 
neglected in $\Lambda_{\rm eff}$ and $\kappa_{\rm eff}$, 
and $\kappa$ is the constant contribution (if any) to 
$\kappa_{\rm eff}$.
Thus, in recent times the
main effect of the gravitational renormalization terms is to 
cause a large finite renormalization of the gravitational constant.
This same renormalization of the gravitational constant
also follows from the gravitational field equation
corresponding\cite{er2} to Eq.(4.2) of Ref.\cite{pt1}.
The value of Newton's gravitational constant $G$ in recent times
(and in fact going back to times when the magnitude of $R$ first
became small with respect to $M^2$ for the most massive particles)
satisfies
\begin{equation}
(16\pi G)^{-1} = (\kappa - 
 \sum_i A_i B_i N_i M_i^2),
\label{eq:Newton1}
\end{equation}
where the summation has been restored.

It is conceivable that independent limits on the value of the constant
$\kappa$ may come from considering the evolution of the very early
universe. However, a plausible conjecture may be made at this time.
If, as expected, there
are massive particles having masses that are at the GUT scale
or a significant fraction of the Planck scale, then
the sum in Eq.(\ref{eq:Newton1}) is sufficiently large
that it may account for the full value of Newton's constant.
If we assume that this is the case, i.e., that $\kappa$ is
negligible compared to the sum,
then we obtain a testable relation between the masses
and multiplicities of the most massive particles 
and the value of Newton's constant $G$. Writing 
$G = M_{\rm Pl}{}^{-2}$, where $M_{\rm Pl} = 1.2\times 10^{19}$ GeV,
this relation is
\begin{equation}
(16\pi) \sum_i A_i B_i N_i (M_i/M_{\rm Pl})^2 = -1,
\label{eq:mass1}
\end{equation}
where the summation is over each type of 
particle that, in the present universe, satisfies the 
condition $|B_i R/ M_i{}^2| \ll 1$.
If the sum is dominated by
$N$ scalar bosons of mass $M$, then $A = (64\pi^2)^{-1}$,
and if we assume that $B \approx -1$, then Eq.(\ref{eq:mass1})
would imply that
\begin{equation}
(4 \pi)^{-1} N (M/M_{Pl})^2 \approx 1.
\label{eq:estimate1}
\end{equation}
This appears to be a reasonable approximate relation in the light 
of current theories that predict multiplets
of elementary particles having very large masses.
With $B \approx -1$, the corresponding terms in the sum in 
Eq.(\ref{eq:lambdasum}) will give a positive contribution to the
effective cosmological constant in the early universe, favoring
early inflation.

In summary, the conjecture that in recent times
the coefficient of the scalar curvature $R$
in the effective action is dominated
by the summation term in Eq.(\ref{eq:Newton1}) gives
Eq.(\ref{eq:mass1}), which can immediately be tested against 
theories of elementary particles and strings.
At the level of quantum field theory in curved spacetime, this
conjecture is a new embodiment of Sakharov's idea of induced 
gravity.\cite{sak,adl} In a fully unified finite theory, it may be 
hoped that a relation of the type of Eq(\ref{eq:mass1}) will
follow entirely from the theory, without the need
for subsidiary assumptions such as the smallness of the constant
$\kappa$ relative to the sum in Eq.(\ref{eq:Newton1}).
Nevertheless, the relation of Eq.(\ref{eq:mass1}) between
Newton's constant and the elementary particle masses is a
fairly plausible and testable consequence of quantum field 
theory and general relativity.

I thank Dr.~Sean Carroll, Dr.~Andrew Liddle,
and Dr.~Robert Myers for very helpful and perceptive
comments on the  first draft of this paper, bringing
my attention to the renormalization of the gravitational
constant. I thank the National Science Foundation for
support under grant PHY-9507740.


\begin{thebibliography}{1}
\bibitem{nl}
By non-perturbative or non-local, we mean terms which cannot
be expressed as any finite polynomial with constant coefficients
in the metric and its derivatives.
\bibitem{er1}
The minus sign in Eq.(4.3) of Ref.\cite{pt1} should be plus, as here.
\bibitem{er2}
In Ref.\cite{pt1}, Eq.(4.4) should include a contribution from the
second $R_{\mu\nu}$ term of Eq.(4.2).
\end{thebibliography}
\end{document}